\documentclass{emulateapj}
\usepackage{amsmath, amsthm, amssymb}

\slugcomment{Submitted to ApJ}

\shorttitle{BLACK HOLE MASSES AND STELLAR VELOCITY DISPERSION}
\shortauthors{GASKELL}

\begin{document}

\title{AN IMPROVED [O\,III] LINE WIDTH TO STELLAR VELOCITY DISPERSION CALIBRATION: CURVATURE, SCATTER, AND LACK OF EVOLUTION IN THE BLACK-HOLE MASS VERSUS STELLAR VELOCITY DISPERSION RELATIONSHIP}

\author{C. MARTIN GASKELL}

\affil{Department of Astronomy, University of Texas, Austin, TX 78712-0259.}

\email{gaskell@astro.as.utexas.edu}

\begin{abstract}

An improved transformation of the full width at half maxima (FWHM) of the [O\,III] $\lambda$5007 line in AGNs to the
stellar velocity dispersion, $\sigma_*$, of the host galaxy is given. This significantly reduces the systematic errors in using
FWHM({[O\,III]}) as a proxy for $\sigma_*$. AGN black hole masses, $M_\bullet$, estimated using the Dibai single epoch spectrum method,
are combined with the new estimates of $\sigma_*$ to give a revised AGN $M_\bullet$\,--\,$\sigma_*$ relationship extending up to high masses.  This shows that for the most massive black holes, $M_\bullet$ is systematically higher than predicted by extrapolation of $M_\bullet \propto \sigma_*^4$ to high masses.  This supports recent suggestions that stellar dynamical masses of the most massive black holes have been systematically underestimated. The steepening of the $M_\bullet$\,--\,$\sigma_*$ relationship is consistent with the absence of very high $\sigma_*$ galaxies in the local universe and with the curvature of the Faber-Jackson relationship.  There appears to be significantly less intrinsic scatter in the $M_\bullet$\,--\,$\sigma_*$ relationship for galaxies with $M_\bullet > 10^9 M_\sun$.  It is speculated that this is connected with the core elliptical versus extra-light elliptical dichotomy.  The low scatter in the high end of the $M_\bullet$\,--\,$\sigma_*$ relationship implies that the transformation proposed here and the Dibai method are good indicators of $\sigma_*$ and $M_\bullet$ respectively.  There is no evidence for evolution of the $M_\bullet$\,--\,$\sigma_*$ relationship over time.

\end{abstract}


\section{INTRODUCTION}

There is considerable interest in the correlation of the masses,
$M_\bullet$, of supermassive black holes (SMBHs) with the stellar
velocity dispersion, $\sigma_*$, of the bulge of the host galaxy
\citep{gebhardt+00,ferrarese+merritt00,kormendy+gebhardt01} because of its important
implications for the formation and evolution of galaxies and their
SMBHs. It is widely recognized that for understanding the origin of the
$M_\bullet$\,--\,$\sigma_*$ relationship it is particularly important to
know the masses of black holes in the most massive and least massive
galaxies, and to study the evolution of the $M_\bullet$\,--\,$\sigma_*$
relationship over cosmic time. Having the correct form of the
relationship is also important for
constructing the black hole mass function in the local universe and
reconciling this with the luminosity function of AGNs, energy-generation efficiencies, relative accretion
rates (Eddington ratios), AGN duty cycles, and the evolution of all
these quantities over the history of the universe.  The predictions from the bulge luminosity, $L_{bulge}$, and from the
$M_\bullet$\,--\,$\sigma_*$ relationship conflict for the
highest-luminosity galaxies because the $M_\bullet$\,--\,$L_{bulge}$
relationship predicts that the brightest galaxies have
$M_\bullet \sim 10^{10} M_{\sun}$, while the $M_\bullet$\,--\,$\sigma_*$
relationship predicts $M_\bullet < 3 \times 10^9 M_{\sun}$ at all
times \citep{bernardi+07b,lauer+07}.  The number of high-mass SMBHs predicted by
the $M_\bullet$\,--\,$\sigma_*$ relationship is also in conflict with
the number of relic high-mass SMBHs predicted by the volume density
of the most luminous AGNs \citep{lauer+07}.

By far the easiest way to estimate black hole masses is to use
type-1 (near face-on) AGNs.  To do this one needs the velocity dispersion
of the broad-line region (BLR) gas, which can be readily obtained
from the H$\beta$ line width, and an effective radius. It is also important to establish that
the motions of the gas being used are dominated by gravity.  The most direct way of
determining an effective radius is via ``reverberation mapping'' using light echoes following AGN variability
\citep{lyutyi+cherepashchuk72,cherepashchuk+lyutyi73}.  This is most readily
accomplished by cross correlating line and continuum variability \citep{gaskell+sparke86}.
Velocity-resolved light echoes have shown that BLR gas motions are
gravitationally dominated (\citealt{gaskell88,koratkar+gaskell89}; see \citealt{gaskell+goosmann08}
and \citealt{gaskell09} for reviews of the evidence for this) and thus permit estimation of $M_\bullet$ from the BLR. Since then,
$M_\bullet$ has been estimated from reverberation mapping for several dozen AGNs (see \citealt{vestergaard+peterson06} for a
recent compilation).  An external check on the accuracy of reverberation mapping masses
is provided by the tightness of the correlations of the  $M_\bullet$ estimates with $\sigma_*$ and $L_{bulge}$.  The bulge luminosities obtained by \citet{bentz+09a} imply that the error in $\log M_\bullet$ determined by reverberation mapping is less than $\pm 0.33$ dex.

Reverberation mapping is unfortunately very labor intensive, but \citet{dibai77} argued that BLR radii,
and hence masses, could be inferred indirectly from photoionization considerations from single-epoch spectra.  Mass estimates from a large number of spectra of NGC\,5548 in high and low states \citep{denney+09} show that random and systematic measuring errors in $\log M_\bullet$ determined by the Dibai method are small ($\approx 0.11$ dex).  \citet{bochkarev+gaskell09} have shown that Dibai's original mass estimates \citep{dibai77,dibai80,dibai84} are in good agreement with reverberation-mapping mass estimates. The same is true for more recent mass determinations by the Dibai method \citep{vestergaard+peterson06}.  As with reverberation-mapping-determined masses, an independent external check is provided by correlations with the properties of the host-galaxy bulges. \citet{gaskell+kormendy09} find a scatter of $\pm 0.23$ in $\log M_\bullet$. These studies suggest that, with care, the Dibai method, which is about two orders of magnitude less labor intensive than reverberation mapping, gives $M_\bullet$ accurate to $\lesssim \pm 0.25$ dex.

A problem in investigating the $M_\bullet$\,--\,$\sigma_*$ relationship for AGNs is that strong optical continuum continuum emission from the
AGN accretion disk can make measuring $\sigma_*$ difficult. This is exacerbated as one goes to higher redshifts as the most
readily observable stellar absorption lines move into the IR.  \citet{nelson00} proposed using
FWHM$_{[O\,III]}$/2.35 as a proxy for $\sigma_*$ because the [O\,III] lines are strong and readily observable.  For a Gaussian
velocity dispersion, $\sigma = $FWHM$/2.35$. By using FWHM$_{[O\,III]}$/2.35 Nelson showed that active galaxies also obey
the $M_\bullet$\,--\,$\sigma_*$ relationship. The use of FWHM$_{[O\,III]}$ as a proxy for $\sigma_*$ has subsequently been widely
adopted because the [O\,III] lines are readily observable and they allow studies of the $M_\bullet$\,--\,$\sigma_*$ relationship to be
extended to higher redshifts and higher luminosities \citep{shields+03}.

An obvious concern with using FWHM$_{[O\,III]}$ as a proxy for $\sigma_*$ is that gas, unlike stars, is highly susceptible to
non-gravitational forces and dissipation, and it has long been known that the motions of the gas in galaxies are complicated.  In this paper I show that using FWHM$_{[O\,III]}$ as a proxy for $\sigma_*$ {\em underestimates} $\sigma_*$ for narrow [O\,III] lines and {\em overestimates} it for
broad [O\,III] lines.  I give a simple prescription for removing this bias, and discuss the implications for the $M_\bullet$\,--\,$\sigma_*$ relationship.

\section{USING THE FWHM OF [O\,III] AS A PROXY FOR STELLAR VELOCITY DISPERSION}

The justification of \citet{nelson00} in using FWHM$_{[O\,III]}$/2.35
as a proxy for $\sigma_*$ was that in the compilation of
\citet{nelson+whittle95} and \citet{nelson+whittle96}, the {\em
average} FWHM$_{[O\,III]}$/2.35 was in good agreement with their average
directly measured $\sigma_*$.   A number of other studies (e.g.,
\citealt{boroson03,onken+04,greene+ho05,bonning+05}) have supported
the use of FWHM$_{[O\,III]}$/2.35 as a proxy for $\sigma_*$,
although one with significant scatter. This scatter is not
surprising as it has long been known \citep{burbidge+59} that the
narrow emission lines are blueshifted and this has been shown in a
number of cases to be the result of outflow. The widths of the narrow-line region (NLR)
lines also depend on the ionization potential and critical densities
of the lines \citep{derobertis+osterbrock86}, so different lines
will give systematically different FWHMs.  With these complications we therefore expect a
merely empirical correlation between FWHM$_{[O\,III]}$/2.35 and $\sigma_*$ at best.

Fig.\@ 1 shows FWHM$_{[O\,III]}$/2.35 and measured $\sigma_*$ values from
\citet{nelson+whittle95} and \citet{nelson+whittle96}. Additional or updated $\sigma_*$ measurements
were taken from \citet{nelson00} (Mrk 79, Mrk 110, Mrk 279),
\citet{nelson+04} (NGC 4151 and NGC 5548), and \citet{onken+04} (NGC 7469, Mrk 817, Ark 120).
Updated FWHM$_{[O\,III]}$ measurements were taken from \citet{nelson00} for NGC 7469, Mrk 79, Mrk 110, Mrk 279, Mrk 817, and Ark 120.
Strong radio sources (objects in Table 7 of \citealt{nelson+whittle95}) are indicated by the addition of a ``+'' to type-1 objects and a ``$\times$'' to type-2 objects.

It can be seen from Fig.\@ 1 that while FWHM$_{[O\,III]}$/2.35 for all AGNs agrees on average with $\sigma_*$ for the sample as a whole, there is substantial scatter.    \citet{nelson+whittle96} find an rms scatter of $\pm 0.20$ dex (see their Fig.\@ 7a).  The rms scatters for both the type-1 and type-2 AGNs in Fig.\@ 1 are similar ($\pm 0.20$ dex $\pm 0.19$ dex respectively).  The rms of the quoted measuring errors of $\sigma_*$ is $\pm 0.07$ dex.  Comparison of FWHMs of [O\,III] with the FWHMs of similar ionization [S III] lines (Table 3 of \citealt{nelson+whittle95}) suggests that the rms error in the measurement of each line is $\pm 0.06$ dex.  The expected scatter in Fig.\@ 1 from measuring errors alone is thus expected to be $\approx \pm 0.10$ dex.

\begin{figure}
\vspace*{0.3cm}
\centering
\includegraphics[width=70mm]{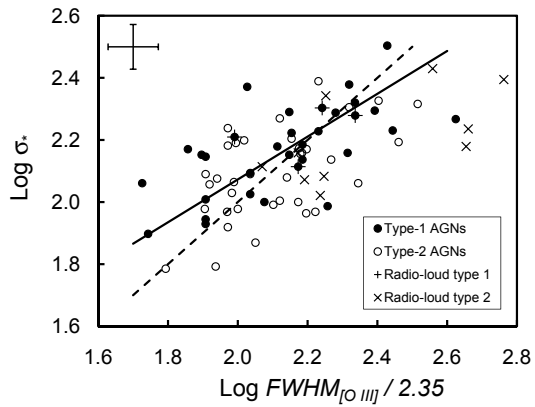} 
\caption{Measured stellar velocity dispersion, $\sigma_*$, as a
function of FWHM$_{[O\,III]}$/2.35 for AGNs.  The dotted line is the equation
$\sigma_* =$ FWHM$_{[O\,III]}$/2.35 and the solid line is an OLS-bisector fit \citep{isobe+90} for the type-1 AGNs only.} \label{all_data}
\end{figure}

In addition to the large scatter in Fig.\@ 1, it can be see that there is a {\em systematic} deviation from $\sigma_*$ = FWHM$_{[O\,III]}$/2.35 for wide [O\,III] lines as has already been noted by \citet{nelson+whittle96}.    It is, of course, important for studying the $M_\bullet$\,--\,$\sigma_*$ relationship for AGNs that there be no systematic errors in estimating $\sigma_*$.  \citet{nelson+whittle96} have already pointed out that the systematic deviation is strongest for radio-loud AGNs.  Fig.\@ 1 shows that it is actually the {\em type-2} radio-loud AGNs which show the largest systematic effect.  This argues that the NLR outflows are preferentially near the equatorial plane.  Type-1 AGNs show a similar systematic bias when [O\,III] is used as a proxy for $\sigma_*$.  This is shown by the OLS-bisector fit \citep{isobe+90} to the type-1 AGNs alone in Fig.\@ 1.  The effect is quite substantial.  For type-1 AGNs with FWHM$_{[O\,III]}$/2.35 $< 100$ km s$^{-1}$ the median under prediction of $\sigma_*$ is 0.22 dex.
Since $M_\bullet \propto \sigma_*^4$, this translates into a 0.9 dex systematic error $\log M_\bullet$.

The NLR line widths are consistent with other observations showing that emission-line gas velocities in bulges tend to be sub-virial.  For example, in a large study of elliptical galaxies by \citet{phillips+86} the mean FWHM/2.35/$\sigma_*$ from their [N~II] FWHMs is $0.73 \pm 0.03$.  Spatially resolved {\it Hubble Space Telescope} spectroscopy of NLRs shows that the innermost gas is commonly moving more slowly than gas further out \citep{rice+06}.

Sub-virial NLR motion is obviously being seen because gas is settling into a flattened distribution.  Since type-1 AGNs are viewed close to face-on, this will reduce the observed FWHM for gas in a flattened distribution.  The smaller systematic offset for type-2 AGNs would be because they are viewed more obliquely.  It can be seen in Fig.\@ 1 that both the overestimation of $\sigma_*$ and the difference between type-1 and type-2 AGNs decreases as the FWHM gets wider. The reason for the decreasing difference between type-1 and type-2 AGNs with increasing FWHM could be that for  type-1 AGNs with the largest FWHMs the torus opening angle is larger than average, and so they can be seen at higher inclinations.  The extra width of [O\,III] could also be associated with an outflow.  This must certainly be going on for the AGNs with the highest [O\,III] FWHMs.

\section{AN IMPROVED RELATIONSHIP BETWEEN [O\,III] FWHM AND $\sigma_*$}

Whatever the cause of the systematic deviations in Fig.\@ 1, it possible to remove the systematic effect empirically by using a non-linear transformation from $\sigma_{[[O\,III]]corr}$ to $\sigma_*$.   For type-1 AGNs a corrected estimate of the velocity dispersion can be defined by the equation:

\begin{align*}
\log \sigma_{[[O\,III]]corr} = (0.67 \pm 0.09) \log [FWHM_{[O\,III]}/2.35] \\ + (0.74 \pm 0.02)~~~~~~(1)
\end{align*}

\noindent where the coefficients have been chosen so that an OLS-bisector fit to $\sigma_*$ and $\sigma_{[O\,III]corr.}$ gives the equation $\sigma_* = \sigma_{[O\,III]corr.}$.  The comparison between the measured $\sigma_*$ values and the estimates using Eq.\@ 1 are shown in Fig.\@ 2.  The scatter for the type-1 AGNs has been reduced to $\pm 0.12$ dex. Type-2 AGNs cannot be used to get $M_\bullet $ via the Dibai method and hence will not be considered further, but the corresponding transformation for type-2 AGNs has a slope of $0.82 \pm 0.11$.  This is consistent with the slope of Eq.\@ 1, but there is an offset of 0.06 dex in the sense that Eq.\@ 1 would slightly over predict $\sigma_*$ for type-2 AGNs.  For radio-quiet type-2 AGNs the scatter is reduced to $\pm 0.12$ dex, and even for radio-loud AGNs it is reduced to $\pm 0.14$ dex.

\begin{figure}
\vspace*{0.3cm}
\centering
\includegraphics[width=70mm]{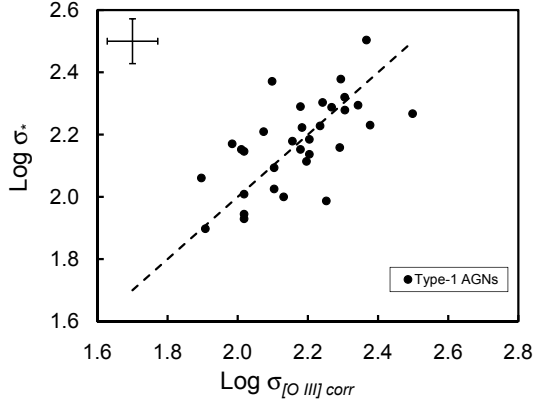} 
\caption{Measured stellar velocity dispersion, $\sigma_*$, as a
function of the corrected [O\,III] line velocity dispersion, $\sigma_{[O\,III]corr}$,
from Eq. 1.  The diagonal line is the equation $\sigma_* = \sigma_{[O\,III]corr.}$} \label{all_data}
\vspace*{0.3cm}
\end{figure}

\section{THE $M_\bullet$\,--\,$\sigma_*$ RELATIONSHIP FOR AGNS}

\citet{shields+03} made estimates of $M_\bullet$ and $\sigma_*$ for a large sample of AGNs with observations by \citet{brotherton96a,brotherton96b},  \citet{grupe+99}, \citet{mcintosh+99}, \citet{dietrich+02}, and themselves.  This sample has an advantage over the nearby AGN sample of \citet{greene+ho06} in having AGNs covering the upper end of the relationship defined by normal galaxies (see \citealt{gultekin+09}).
\citet{shields+03} calculated black hole masses from FWHMs of H$\beta$ by the Dibai method, and $\sigma_*$ was just taken to be FWHM$_{[O\,III]}$/2.35.  Details of the treatment of the data can be found in \citet{shields+03}.  Fig.\@ 3 shows the $M_\bullet$\,--\,$\sigma_*$ relationship for these BLR blackhole mass estimates and also for the non-BLR blackhole mass estimates compiled by  \citet{gultekin+09}.  For three galaxies revised masses are also shown (see discussion below).  $M_\bullet$ estimates from BLR kinematics have always been normalized to the stellar-dynamical $M_\bullet$\,--\,$\sigma_*$ relationship.  The \citet{shields+03} masses have been renormalized to the updated \citet{gultekin+09} relationship.  The renormalization has been made over the range $2.0 < \log \sigma_{[[O\,III]]corr} < 2.4$ where the corrections to the [O\,III] estimates of $\sigma_*$ are smallest (see Fig.\@ 1).  The \citet{shields+03} masses need to be decreased by 0.11 dex to minimize their median residual from the \citet{gultekin+09} line in this range.  This small renormalization has a negligible effect on the results.

\begin{figure}
\vspace*{0.3cm}
\centering
\includegraphics[width=70mm]{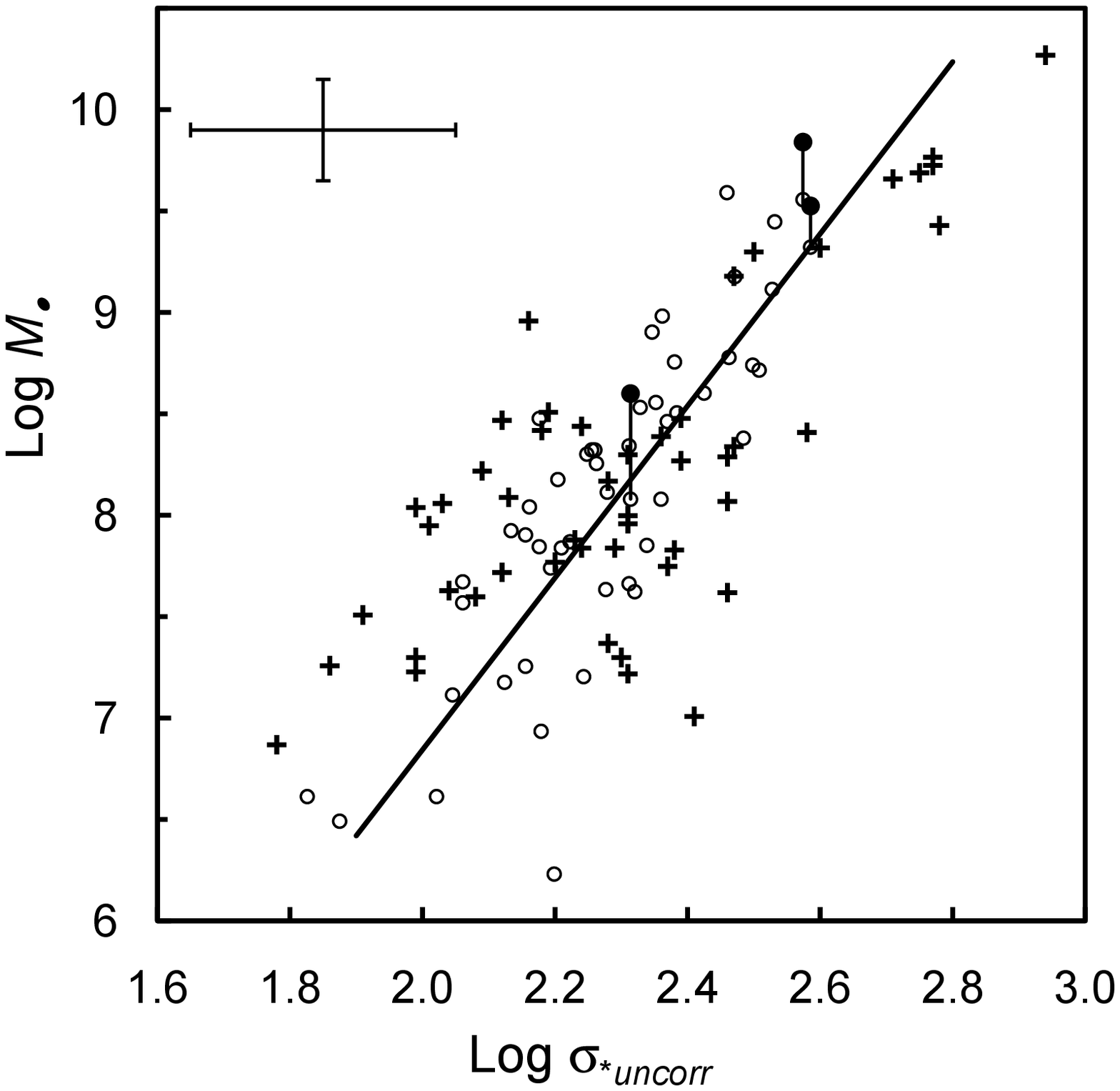} 
\caption{The $M_\bullet$\,--\,$\sigma_*$ relationship for the radio-quiet AGNs of \citet{shields+03} (shown as crosses) and for the non-BLR mass estimates compiled by \citet{gultekin+09} (shown as open circles).  For the \citet{shields+03} AGNs $\sigma_*$ has been taken simply to be FWHM$_{[O\,III]}$/2.35 with no correction.  The solid circles joined with short vertical lines to the open circles below them represent the $M_\bullet$ revisions of \citet{humphrey+08}, \citet{gebhardt+thomas09}, and \citet{vandenbosch08}.  A typical error bar for the AGNs is shown at the upper left.  The error in $\sigma_{*uncorr}$ is the scatter found by \citet{nelson+whittle95}, and the error in $M_\bullet$ has been taken to be $\pm 0.25$ following \citet{bochkarev+gaskell09}. The diagonal line is the \citet{gultekin+09} fit. } \label{all_data}
\end{figure}

It can be seen from Fig.\@ 3 that the \citet{shields+03} objects deviate systematically from the relationship given by the \citet{gultekin+09} galaxies.  For $\log \sigma_{*uncorr} < 2.2$ the inferred masses of the \citet{shields+03} galaxies all lie above the \citet{gultekin+09} fit, and for $\log \sigma_{*uncorr} > 2.7$ they all lie below.

Fig.\@ 4 plots the \citet{shields+03} galaxies with the corrected estimates of the stellar velocity dispersions given by Eq.\@ 1.  It can be seen that the scatter is less and the systematic differences from the \citet{gultekin+09} points are reduced.  The agreement is particularly good at the high-mass end.  The typical error bar (upper left corner) is now smaller.  The uncertainty in the slope of Eq.\@ 1 introduces an additional uncertainty in $\sigma_*$ for the highest and lowest mass black holes.  The amplitude of this uncertainty for a typical high-mass black hole is indicated by the horizontal error bar at the upper right.  One of the free parameters in the Dibai method is the slope of the relationship between the effective radius, $R$, of the BLR and the optical luminosity, $L_{AGN}$.  This introduces an additional uncertainty in the $M_\bullet$\,--\,$\sigma_*$ relationship for high- and low-mass AGNs.  \citet{shields+03} took $R \propto L^{0.5}$.  This is in good agreement with the best current estimate of $R \propto L^{0.52}$ \citep{bentz+09b}.

\begin{figure}
\vspace*{0.3cm}
\centering
\includegraphics[width=70mm]{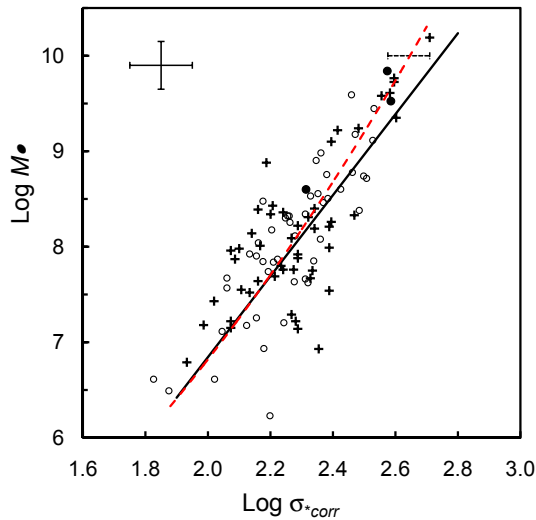} 
\caption{The $M_\bullet$\,--\,$\sigma_*$ relationship for the BLR and non-BLR mass estimate samples as in Fig.\@ 3 but with correction of the stellar
velocity dispersion estimates for the AGNs as described in the text.  For clarity the three older mass estimates in Fig.\@ 1 that have new values have been omitted.  The dashed red curve is Eq.\@ 2.  The dashed horizontal error bar in the upper right indicates the effect of the uncertainty in the slope in Eq. 1. All other symbols have the same meaning as in Fig.\@ 3.  The error in $\log \sigma_{*corr}$ has been taken to be the scatter in Fig.\@ 2.} \label{all_data}
\end{figure}

\section{DISCUSSION}

\subsection{The masses of the most massive black holes}

The revised AGN $M_\bullet$\,--\,$\sigma_*$ relationship of Fig.\@ 4 supports recent upwards revisions of stellar dynamical $M_\bullet$ estimates in massive galaxies.  Two effects are responsible for the revisions.  The first is that including dark halos in stellar dynamical modelling of the most massive galaxies (see \citealt{gebhardt+thomas09}) increases the estimated masses of their black holes by about a factor of two.  The masses of two of the most massive black holes have been increased already (see Fig.\@ 3), and more masses of the most massive black holes are expected to increase similarly (K. Gebhardt - private communication).  The second effect is that triaxial orbit calculations can give a higher $M_\bullet$ than axisymmetric models.  This has been demonstrated for NGC~3379 by \citet{vandenbosch08} and \citet{vandenbosch+09}.  This is likely to be a general effect and triaxial galaxies are believed to dominate among high-mass galaxies.  The two systematic effects are independent and are expected to act together for some high-mass galaxies.  Further upward revisions of the masses of the most massive black holes are thus likely.  Some support for this already comes from direct measurement of $\sigma_*$ in AGNs with black hole masses determined from reverberation mapping.  For some of the AGNs with $\log \sigma_* \sim 2.3$ their masses can be up to $\sim 0.7$ dex above the stellar dynamical $M_\bullet$\,--\,$\sigma_*$ relationship \citep{dasyra+07,watson+08,woo+08}.

\subsection{Curvature in the $M_\bullet$\,--\,$\sigma_*$ relationship}

It can be seen from Fig.\@ 4 that for both the \citet{gultekin+09} galaxies {\em and} the \citet{shields+03} AGNs, the G\"ultekin et al.\@ fit underpredicts $\log M_\bullet$ by about 0.5 dex for galaxies with $\log M_\bullet > 9$.  Even before the recent upwards revision of mass estimates of the highest mass black holes (see previous section), \citet{wyithe06} had argued that a curved log-quadratic relationship was a better fit to the $M_\bullet$\,--\,$\sigma_*$ relationship than a simple power-law.  \citet{hu08} also suggests upwards curvature of the relationship at the high-mass end.

It is now recognized that the shape of the $M_\bullet$\,--\,$\sigma_*$ relationship depends on the type of galaxy being looked at \citep{graham08,hu08}.  Defining the form of the relationship for a mixed sample of galaxies is particularly problematic at the lower end.  It will take further work to establish the true shape of the top of the $M_\bullet$\,--\,$\sigma_*$ relationship for AGNs, but if we assume that the \citet{gultekin+09} line is a good representation for $\log \sigma_* \lesssim 2.5$, then the median slope of a power law from $\log \sigma_* = 2.5$ to the top ten points in Fig.\@ 4 is 5.5.  For computational convenience in interpreting observations, these two lines can be approximated by the log-quadratic:

\begin{equation*}
\log M_\bullet= 1.1 (\log \sigma_*)^2 - 0.2 \log \sigma_* + 2.82~~~~~~~~(2)
\end{equation*}

\noindent It should be noted that, for the points in Fig.\@ 4, the reduction in $\chi^2$ per degree of freedom in going from a power-law to a log-quadratic relationship is not significant.  \citet{gultekin+09} get a similar result for their galaxies alone.  This is because of the scatter in the lower half of the relationship.  Whether or not a log-quadratic curve is formally needed depends on the choice of low $\sigma_*$ galaxies included.  The significant thing about Fig.\@ 4, however, is that the slope of the $M_\bullet$\,--\,$\sigma_*$ relationship {\em at the high-mass end} is steeper than in the \citet{gultekin+09} fit and similar previous fits (e.g., \citealt{tremaine+02}).

Since the determination of the slope of the upper part of the $M_\bullet$\,--\,$\sigma_*$ relationship (see above) depends in part on stellar dynamical mass estimates, some of which are expected to increase (see section 5.2), the slope of the top of the relationship could well increase further. Even though curvature in the $M_\bullet$\,--\,$\sigma_*$ relationship is currently only marginally statistically significant (see discussion in \citealt{gultekin+09}), there are good reasons to believe that the relationship {\em must} curve upwards strongly.  This is because it has long been known \citep{oegerle+hoessel91} that the relationship between galaxy luminosity and $\sigma_*$ \citep{faber+jackson76} curves strongly at the high-luminosity end and $\sigma_*$ saturates. The highest velocity dispersion galaxies found by \citep{oegerle+hoessel91} have $\log \sigma_* \sim 2.6$, and further searches have failed to turn up any galaxies with $\log \sigma_* > 2.65$ \citep{salviander+08}.  The velocity dispersions for the most massive black holes in Fig.\@ 4 are consistent with this limit.  The steepening in Fig.\@ 4 removes the discrepancy between the $M_\bullet$\,--\,$\sigma_*$ and $M_\bullet$\,--\,$L_{bulge}$ relationships \citep{bernardi+07b,lauer+07}, and as \citet{wyithe06} points out, the steepening increases the local density of the highest mass black holes.

\subsection{Dispersion in the $M_\bullet$\,--\,$\sigma_*$ relationship as a function of $M_\bullet$}

Another interesting thing in Fig.\@ 4 is the tightness of the $M_\bullet$\,--\,$\sigma_*$ relationship at the high-mass end.  This impression is confirmed in Fig.\@ 5 which shows the dispersion in $\log \sigma_*$ for the combined samples in Fig.\@ 4 as a function of $\log M_\bullet$.  It can be seen that the dispersion for the highest-mass objects is significantly lower than for the remaining objects.  An F-test gives a probability, $p = 0.01$ that the dispersions of objects with $M_\bullet > 10^9$ and $10^{7.5} < M_\bullet < 10^9 M_\sun$ are drawn from the same parent population.  The decrease in dispersion at high masses is seen for the BLR estimates and the \citealt{gultekin+09} estimates separately, but at lower significance.

\begin{figure}[t!]
\vspace*{0.8cm}
\centering
\includegraphics[height=65mm]{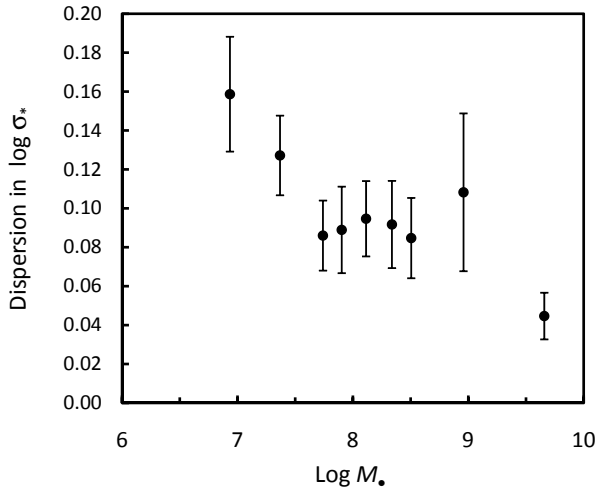} 
\caption{The dispersion $\log \sigma_*$ (in standard deviations) as a function of $\log M_\bullet$.  The bins have been chosen to have equal numbers of objects.} \label{all_data}
\end{figure}

The scatter at the lower end of the $M_\bullet$\,--\,$\sigma_*$ relationship is similar to that found by \citet{greene+ho06} from direct measurements of the stellar velocity dispersion by for AGNs with BLR masses determined by the Dibai method.  It is also similar to that found by \citet{onken+04} for reverberation-mapped AGNs with a mixture of direct measurements of stellar velocity dispersions and estimates from FWHM$_{[O\,III]}$/2.35).  In Fig.\@ 1 of \citet{greene+ho06} the increase in scatter going from the high-mass galaxies with stellar dynamical determinations to the lower-mass AGNs  is particularly striking.  The results presented here argue that this effect is real and not a consequence of larger measuring errors for the AGNs.

The increase in scatter in the $M_\bullet$\,--\,$\sigma_*$ relationship as one goes to lower black hole and galaxy masses probably tells us something about blackhole growth associated with the different processes that dominate the evolution of galaxies of different masses.  Elliptical galaxies and classical bulges in spirals are the result of violent mergers.  For lower mass galaxies, secular evolution dominates instead.  Secular evolution can be recognized by the presence of bars, pseudobulges, and other features (see \citealt{kormendy+kennicutt04}). It is now know that galaxies with bars and pseudo-bulges have systematically lower blackhole masses than predicted by the $M_\bullet$\,--\,$\sigma_*$ relationship for ellipticals and classical bulges \citep{hu08,graham08}.  Secular evolution must therefore be a cause of some of the increase in scatter in Fig.\@ 5.  One test of whether it is the {\em only} cause of the trend is to look at the dispersion in $\log \sigma_*$ for the pure ellipticals.  Inspection of the ellipticals in the G\"ultekin et al. sample shows that the dispersion rises from $\approx 0.05$ dex at $\log M_\bullet \approx 9.5$ to $\approx 0.11$ at $\log M_\bullet \approx 7.5$.  Unfortunately, the significance of this trend is low ($p = 0.25$) because of the small sample size.  If the dispersion does indeed rise for the lowest mass ellipticals, this would suggest that secular evolution at low masses is {\em not} the only cause of the dependence of dispersion on black hole mass.

For the highest mass galaxies the low dispersion in $M_\bullet$\,--\,$\sigma_*$ is probably associated with the dichotomy between ellipticals with ``cores'' and those with central ``extra light'' \citep{kormendy+09}.  ``Core'' ellipticals have had their cores scoured out by supermassive black hole binaries \citep{ebisuzaki+91} after ``dry'' dissipationless mergers.  ``Extra-light'' ellipticals have had major star formation in their centers following ``wet'' starburst mergers.  Ellipticals with $M_\bullet > 10^9$ are core ellipticals.  \citet{bernardi+07a} have shown that the brightest ellipticals have a very low scatter in the fundamental plane, and \citet{kormendy+bender09} have shown that core ellipticals have a tight correlation between $\sigma_*$ and the stellar light deficit.  All these things and the lower dispersion in the $M_\bullet$\,--\,$\sigma_*$ relationship point to core ellipticals having very uniform properties.

\subsection{Evolution of the $M_\bullet$\,--\,$\sigma_*$ relationship}

The radio-quiet AGNs in the \citet{shields+03} sample go up to a redshift of 3.286.  \citet{shields+03} claimed that deviations from the $M_\bullet$\,--\,$\sigma_*$ relationship as a function of redshift showed that the relationship was unchanged out to $z \approx 3$ and hence consistent with the growth of massive bulges and black holes occurring simultaneously.  The results presented here support this conclusion even though the analyzes are different.  \citet{shields+03} used uncorrected estimates of $\sigma_*$ (see Fig.\@ 3) which, as discussed above, are systematically too high for the most massive AGNs.  To look for evolutionary effects \citet{shields+03} looked for a redshift dependence in residuals from the \citet{tremaine+02}  $M_\bullet$\,--\,$\sigma_*$ relationship.  The  slope of this relationship is somewhat flatter than the \citet{gultekin+09} slope (4.02 versus 4.24) and significantly flatter than the slope of 5.5 for the upper end of the relationship in Fig.\@ 4.  The flatter slope \citet{shields+03} adopted for the $M_\bullet$\,--\,$\sigma_*$ relationship compensated for the larger estimates of $\sigma_*$ at high masses.

Although the overall $M_\bullet$\,--\,$\sigma_*$ relationship seems similar at low and high redshifts much work needs to be done.  The relationship really needs to be compared, not for heterogeneous collections of galaxies, but for different evolutionary stages of galaxies that will end up similar.  Although there does not seem to be strong evolution of the very highest mass galaxies and their black holes, strong evolution of lower mass galaxies and black holes is quite possible \citep{woo+08}, since there is more scatter in the $M_\bullet$\,--\,$\sigma_*$ relationship.  In studying evolution, however, care needs to be taken to allow for selection biases (see for example \citealt{salviander+07}).

\section{Conclusions}

An improved transformation of FWHM$_{[O\,III]}$ to $\sigma_*$ has been proposed.  This makes the velocity dispersions of the highest mass AGNs consistent with the upper limit on $\sigma_*$ in the nearby universe and requires a steepening of the $M_\bullet$\,--\,$\sigma_*$ relationship for the most massive galaxies. The dispersion about the $M_\bullet$\,--\,$\sigma_*$ relationship decreases as the mass of black holes increases.  The tightness of the relationship for the most massive AGNs provides strong support both for the reliability of the proposed FWHM$_{[O\,III]}$ to $\sigma_*$ transformation and of the Dibai method of estimating black hole masses from single-epoch spectra of type-1 AGNs.  There is no evidence yet evolution of the $M_\bullet$\,--\,$\sigma_*$ relationship from $z \sim 3$ to the present day but much further work on this is needed with better-defined samples.

\acknowledgments

I would like to thank John Kormendy for major encouragement during the course of this investigation, and Sandy Faber, Karl Gebhardt, Jenny Greene, Phil Hopkins, Sarah Salviander, Greg Shields, Remco van den Bosch, and Bev Wills for helpful discussions. This research has been
supported in part by the US National Science Foundation through grant AST 08-03883.

\end{document}